\documentclass[pra,reprint,twocolumn,amsmath,amssymb,superscriptaddress]{revtex4-1}
\usepackage[dvipsnames]{xcolor}

\usepackage{graphicx}
\usepackage{epsfig}
\usepackage{bm}
\usepackage[normalem]{ulem}
\usepackage{siunitx}
\usepackage{soul}

\newcommand{\bra}[1]{\ensuremath{\left\langle{#1}\right\vert}}
\newcommand{\braket}[1]{\ensuremath{\left\langle{#1}\right\rangle}}
\newcommand{\ket}[1]{\ensuremath{\left|{#1}\right\rangle}}

\def\bea{\begin{eqnarray}}
\def\eea{\end{eqnarray}}

\usepackage{sansmath}                     
\sansmath  

\begin{document}

\title{Full Microscopic Simulations Uncover Persistent Quantum Effects in Primary Photosynthesis}

\author{Nicola Lorenzoni}
\affiliation{Institut f\"ur Theoretische Physik und IQST, Albert-Einstein-Allee 11, Universit\"at Ulm, D-89081 Ulm, Germany}
\author{Thibaut Lacroix}
\affiliation{Institut f\"ur Theoretische Physik und IQST, Albert-Einstein-Allee 11, Universit\"at Ulm, D-89081 Ulm, Germany}
\author{James Lim}
\affiliation{Institut f\"ur Theoretische Physik und IQST, Albert-Einstein-Allee 11, Universit\"at Ulm, D-89081 Ulm, Germany}
\author{Dario Tamascelli}
\affiliation{Institut f\"ur Theoretische Physik und IQST, Albert-Einstein-Allee 11, Universit\"at Ulm, D-89081 Ulm, Germany}
\affiliation{Dipartimento di Fisica ``Aldo Pontremoli'', Universit{\`a} degli Studi di Milano, Via Celoria 16, 20133 Milano-Italy}
\author{Susana F. Huelga}\email{susana.huelga@uni-ulm.de}
\affiliation{Institut f\"ur Theoretische Physik und IQST, Albert-Einstein-Allee 11, Universit\"at Ulm, D-89081 Ulm, Germany}
\author{Martin B. Plenio}\email{martin.plenio@uni-ulm.de}
\affiliation{Institut f\"ur Theoretische Physik und IQST, Albert-Einstein-Allee 11, Universit\"at Ulm, D-89081 Ulm, Germany}

\begin{abstract}
The presence of quantum effects in photosynthetic excitation energy transfer has been intensely debated over the past decade. Nonlinear spectroscopy cannot unambiguously distinguish coherent electronic dynamics from underdamped vibrational motion, and rigorous numerical simulations of realistic microscopic models have been intractable. Experimental studies supported by approximate numerical treatments that severely coarse-grain the vibrational environment have claimed the absence of long-lived quantum effects. Here, we report the first non-perturbative, accurate microscopic model simulations of the Fenna-Matthews-Olson photosynthetic complex and demonstrate the presence of long-lived excitonic coherences at 77 K and room temperature, which persist on picosecond time scales, similar to those of excitation energy transfer. Furthermore, we show that full microscopic simulations of nonlinear optical spectra are essential for identifying experimental evidence of quantum effects in photosynthesis, as approximate theoretical methods can misinterpret experimental data and potentially overlook quantum phenomena.
\end{abstract}

\maketitle

\textbf{Introduction}

Photosynthesis is a primary biological process that converts light energy into chemical energy, playing a crucial role in sustaining life on Earth. With advancements in experimental and theoretical techniques, the structure-function relationship of photosynthetic pigment-protein complexes (PPCs) has been investigated~\cite{BlankenshipBook2002}. The electronic parameters of pigments, which absorb light and serve as excitation energy transfer (EET) channels, have been estimated using a combination of quantum chemistry methods and experimentally measured linear spectra, such as linear absorption and circular dichroism~\cite{VultoJPCB1998,RengerBJ2006,BuschJPCL2011}. The vibrational modes of PPCs and their interaction with electronic states are considered the primary source of environmental effects in photosynthetic EET. The phonon spectral density, which describes electronic-vibrational (vibronic) coupling strengths as a function of vibrational frequencies, is regarded as a key parameter in determining the presence of quantum effects in photosynthetic EET under ambient conditions. The phonon spectral density has been estimated using first-principles methods~\cite{RengerJPCB2012,CokerJPCL2016,RheePCCP2018,KleinekathoferPR2023} and various spectroscopic techniques, such as fluorescence line narrowing~\cite{RatsepJL2007}.

Although the structure of microscopic models of PPCs is well understood, theoretically verifying or falsifying the presence of quantum effects in PPCs remains a challenging task. This difficulty arises from the comparable strengths of the electronic and vibrational parameters of PPCs, which cannot be described using perturbative theoretical methods~\cite{Caycedo2022}, such as the Lindblad and Redfield equations. Despite recent advancements in non-perturbative methods~\cite{StrathearnNC2018,TamascelliPRL2019,SomozaPRL2019,CygorekNP2022,LorenzoniPRL2024}, their computational costs grow significantly as the phonon spectral density becomes more structured, making numerically exact simulations of full microscopic PPC models intractable. As a result, non-perturbative simulations of PPCs have been carried out using severely coarse-grained phonon spectral densities~\cite{IshizakiPNAS2009,NalbachPRE2011,KreisbeckJPCL2012,BlauPNAS2018}, although these simplified noise models may underestimate quantum effects in PPCs~\cite{Caycedo2022}. In this work, we show that the dissipation-assisted matrix product factorization (DAMPF) method~\cite{SomozaPRL2019}, enhanced by incorporating thermalized spectral densities~\cite{TamascelliPRL2019} and a systematic construction of pseudomode parameters~\cite{LorenzoniPRL2024}, enables non-perturbative simulations of full microscopic PPC models without any approximations (see the SM for details).

Experimentally, the presence of quantum effects in PPCs has been investigated extensively, but the issue remains unresolved. To monitor EET dynamics in PPCs occurring on sub-picosecond timescales, nonlinear spectroscopic techniques, such as two-dimensional electronic spectroscopy (2DES)~\cite{JonasARPC2003,BrixnerJCP2004} with femtosecond laser pulses, have been employed. These methods provide valuable insights into molecular dynamics but have a fundamental limitation in the search for quantum effects in EET. The laser pulses not only create quantum coherences between electronic excited states, associated with EET, but also induce purely vibrational coherences arising from non-equilibrium phonon dynamics in the absence of any electronic excitations, thereby not involving EET~\cite{ButkusCPL2012,TiwariPNAS2013,PlenioJCP2013}. The excited- and ground-state coherences cannot be separated using experimental techniques alone and require detailed theoretical analysis and numerical simulation. However, due to the intractability of non-perturbative simulations of PPCs using methods available at the time, the interpretation of experimental 2D spectra has relied on approximate theoretical approaches~\cite{DuanPNAS2017,ThyrhaugNC2018}, whose failure will be evaluated in this work.

\begin{figure*}
\includegraphics[width=\textwidth]{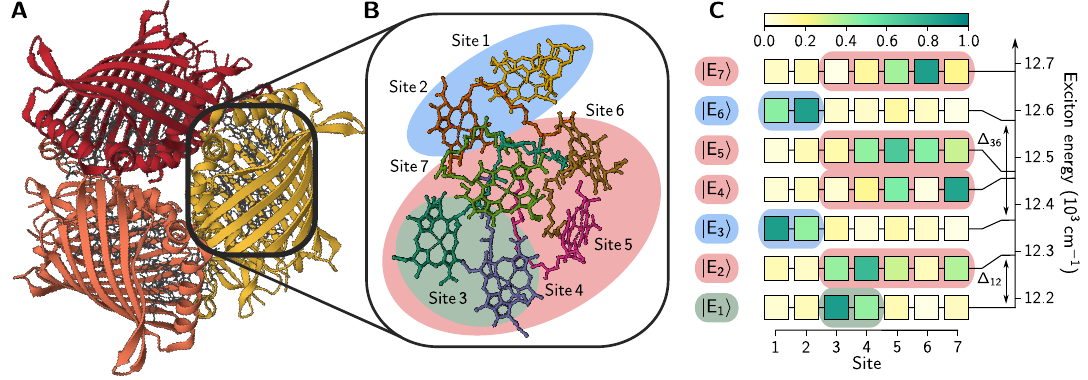}
\caption{\textsf{\textbf{The Fenna-Matthews-Olson complex.} (A) Crystal structure of the trimeric FMO complex from {\it C. Tepidum}~\cite{BermanNucleicAcidsRes2000, TronrudPhotosynthRes2009, FMOstructure, SehnalNucleicAcidsRes2021}. (B) Seven BChl {\it a} pigments in an FMO monomer, labeled by sites 1-7. (C) Population distributions of seven exciton states $\ket{E_k}$ in the site basis, with $\Delta_{jk}=|E_j - E_k|$ representing the energy gap between exciton states. The exciton delocalization is approximately visualized in (B). See the SM for more details.}}
\label{Fig1}
\end{figure*}

The Fenna-Matthews-Olson (FMO) photosynthetic complex from green sulfur bacteria, shown in Fig.~1A, is the first PPC considered in 2D experiments to investigate the presence of quantum effects~\cite{EngelNat2007,EngelPNAS2010,DuanPNAS2017,ThyrhaugNC2018}. In one of the three core experimental studies suggesting the absence of long-lived excited-state coherences in the FMO complex~\cite{CaoSA2020}, the 2D spectra of the FMO complex at $77\,{\rm K}$ were interpreted as follows~\cite{ThyrhaugNC2018}: excited-state coherences decay within $240\,{\rm fs}$ and long-lived coherences persisting on a picosecond timescale have a vibrational origin. However, the approximate theoretical method employed in Ref.~\cite{ThyrhaugNC2018} failed to explain the line shapes of experimentally measured quantum coherences, known as beating maps. In this work, we report the EET dynamics of the FMO complex at $77\,{\rm K}$ and $300\,{\rm K}$, accurately computed for the first time using the full microscopic model, to show that under realistic vibrational environments, excitonic coherences in the FMO complex can persist on a picosecond timescale at both temperatures.

In another 2D study~\cite{DuanPNAS2017}, the quantum coherences of the FMO complex measured at room temperature were attributed to vibrational origin, supported by an approximate theory using a severely coarse-grained phonon spectral density. Ref.~\cite{DuanPNAS2017} has been considered evidence for the absence of quantum effects in the FMO complex under ambient conditions. For the coarse-grained phonon spectral density considered in Ref.~\cite{DuanPNAS2017}, we demonstrate that the oversimplified noise model significantly underestimates the lifetime of excitonic coherences, making them hardly visible at room temperature. Additionally, we report the 2D electronic spectra of model PPCs, non-perturbatively computed for the first time using the actual phonon spectral density of the FMO complex, to demonstrate that the measures in Ref.~\cite{DuanPNAS2017}, which aim to disprove the presence of quantum effects in the FMO complex based on 2D spectra, do not provide conclusive evidence against the existence of quantum effects.

Finally, FMO mutants with modified electronic parameters were studied in broadband pump probe experiments at $77\,{\rm K}$ and room temperature~\cite{MauriNatChem2018}, revealing that the frequencies of long-lived quantum coherences on picosecond timescales remain essentially unchanged by the mutation. As approximate theoretical approaches predict that excitonic coherence frequencies should change as exciton energies are modified, the long-lived quantum coherences observed in the pump probe experiment were interpreted as vibrational in nature. In this work, we show that the frequencies of long-lived excitonic coherences are essentially invariant under variations in electronic parameters, a surprising effect induced by vibronic mixing between exciton states and the entire vibrational environment.

\textbf{Microscopic model}

\begin{figure*}
\includegraphics[width=0.6\textwidth]{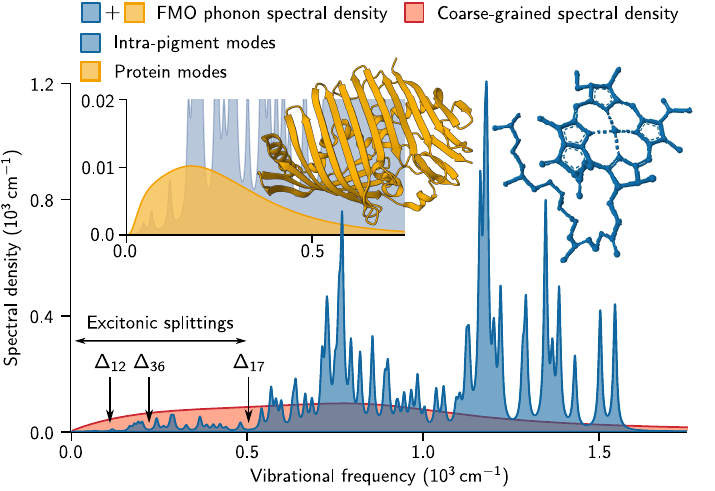}
\caption{\textsf{\textbf{Electron-phonon coupling spectrum.} The vibrational environments of the FMO complex are described by the sum of the phonon spectral density of intra-pigment vibrational normal modes, shown in blue, and that of protein modes, shown in orange. This experimentally measured FMO phonon spectral density~\cite{RatsepJL2007} differs significantly from the coarse-grained phonon spectral density from Ref.~\cite{DuanPNAS2017}, shown in red. The energy differences $\Delta_{ij}=|E_i - E_j|$ between exciton states $\ket{E_i}$ and $\ket{E_j}$ are indicated by black arrows, satisfying $\Delta_{ij} \lesssim 500\,{\rm cm}^{-1}$ (see Fig.~1C). See the SM for more details.}}
\label{Fig2}
\end{figure*}

The FMO complex serves as an EET channel between a large light-absorbing antenna complex, called chlorosome, and a reaction center where charge separation occurs. The FMO complex has a trimeric structure, as shown in Fig.~1A, and an FMO monomer consists of seven or eight pigments, depending on sample preparation, surrounded by a protein scaffold (see Fig.~1B). In molecular ensembles, the FMO complexes typically exhibit finite distributions of electronic parameters due to the static disorder induced by varying local environments. For the mean electronic parameters estimated in Ref.~\cite{RengerBJ2006} (see the SM for details), Fig.~1C summarizes the electronic eigenstates $\ket{E_j}$, called excitons, delocalized over multiple pigments. The lowest energy exciton $\ket{E_1}$ is delocalized over sites 3 and 4, while four exciton states $\ket{E_{2,4,5,7}}$ are delocalized over multiple sites 3-7. The remaining exciton states $\ket{E_3}$ and $\ket{E_6}$ are delocalized over a quasi-dimeric unit of sites 1 and 2.

We note that an accurate description of the phonon spectral density is crucial for estimating electronic parameters from experimentally measured linear optical spectra~\cite{Caycedo2022}. In the main text, we use the electronic parameters from Ref.~\cite{RengerBJ2006}, which have been widely used in FMO studies. In the SM, we show that the long-lived excitonic coherences observed in the main text remain robust under possible refinements of electronic parameters based on numerically exact microscopic simulations of linear optical spectra.

The vibrational environments of the FMO complex are approximately divided into two components~\cite{RatsepJL2007}: intra-pigment and protein modes. The intra-pigment vibrational normal modes exhibit a discrete frequency spectrum, resulting in several tens of narrow peaks in the phonon spectral density, as shown in blue in Fig.~2. The narrow widths indicate a low damping rate of the intra-pigment vibrational modes on a picosecond timescale. In contrast, the protein modes form a quasi-continuous spectrum in the low-frequency region of the phonon spectral density, as shown in orange in Fig.~2. This experimentally estimated phonon spectral density~\cite{RatsepJL2007} is quantitatively similar to those computed using first-principles methods~\cite{CokerJPCL2016,RheePCCP2018,KleinekathoferPR2023}. Additionally, we consider the coarse-grained phonon spectral density from Ref.~\cite{DuanPNAS2017}, shown in red in Fig.~2, which differs significantly from the realistic FMO phonon spectral density.

\textbf{Inter-excitonic coherences}

\begin{figure*}
\includegraphics[width=\textwidth]{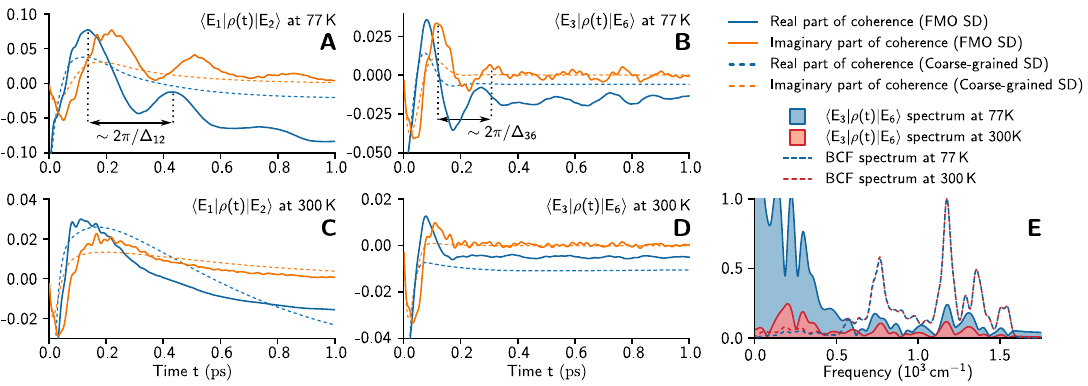}
\caption{\textsf{\textbf{Numerically exact excitonic coherence dynamics.} The real and imaginary parts of excitonic coherences $\langle E_i|\rho(t)|E_j\rangle$ are shown as functions of time $t$, where $\rho(t)$ denotes the reduced electronic density matrix. The results obtained using either the microscopic FMO or the coarse-grained phonon spectral density (SD) are shown in solid and dashed lines, respectively (see Fig.~2). Simulations were performed for the following cases: (A) $(i,j)=(1,2)$ at $77\,{\rm K}$, (B) $(i,j)=(3,6)$ at $77\,{\rm K}$, (C) $(i,j)=(1,2)$ at $300\,{\rm K}$, (D) $(i,j)=(3,6)$ at $300\,{\rm K}$. (E) The frequency spectra of the long-lived excitonic coherence dynamics shown in (B) and (D) over a time window $300\,{\rm fs}\le t \le 1\,{\rm ps}$ are presented, along with the frequency spectra of the bath correlation functions (BCFs) at $77\,{\rm K}$ and $300\,{\rm K}$.}}
\label{Fig3}
\end{figure*}

Using numerically exact non-perturbative methods, we simulated excitonic coherence dynamics in the FMO complex for an initial state where all seven exciton states are superposed with similar amplitudes, a state that is a typical result of laser excitation (see the SM for details). In Fig.~3, we present excitonic coherence dynamics under the influence of either the FMO or the coarse-grained phonon spectral density from Ref.~\cite{DuanPNAS2017} (see Fig.~2).

Fig.~3A shows the coherence dynamics between the two lowest-energy exciton states $\ket{E_1}$ and $\ket{E_2}$ at $77\,{\rm K}$, which has been the primary focus of 2D experiments. For the FMO phonon spectral density, the oscillatory coherence dynamics persist on a picosecond timescale, demonstrating that the lifetime of excitonic coherences at $77\,{\rm K}$ were severely underestimated in Ref.~\cite{ThyrhaugNC2018}. At $77\,{\rm K}$, long-lived coherences were also found for other exciton pairs. As an example, Fig.~3B shows the coherence dynamics between exciton states $\ket{E_3}$ and $\ket{E_6}$. Note that the numerically exact excitonic coherence dynamics exhibit multiple frequencies. While the frequency of the large-amplitude oscillations is close to the energy gap $\Delta_{ij} = |E_i - E_j|$ between the exciton states, high-frequency oscillatory components with relatively small amplitudes are present in both Figs.~3A and B. These high-frequency features disappear when the coarse-grained spectral density is considered.

At room temperature, while the $\Delta_{ij}$-frequency oscillations decay more quickly within $300\,{\rm fs}$, the long-lived high-frequency oscillations observed at $77\,{\rm K}$ are preserved, as shown in Figs.~3C and D. Note that the coarse-grained spectral density completely suppresses all oscillatory features in the excitonic coherence dynamics at $300\,{\rm K}$, significantly underestimating quantum effects in the FMO complex under realistic vibrational environments.

To analyze the nature of the long-lived quantum coherences, Fig.~3E presents the frequency spectrum of the excitonic coherence dynamics between $\ket{E_3}$ and $\ket{E_6}$ over a time window from $300\,{\rm fs}$ to $1\,{\rm ps}$. At both $77\,{\rm K}$ and $300\,{\rm K}$, narrow peaks appear across the entire vibrational frequency range of the intra-pigment modes. A broad peak feature below $500\,{\rm cm}^{-1}$ is observed only at $77\,{\rm K}$, where $\Delta_{36}$-frequency oscillations persist beyond $300\,{\rm fs}$, unlike at room temperature. We note that the narrow-peak structure is similar to that of the frequency spectrum of the bath correlation function (BCF)~\cite{Petruccione2002}, describing the influence of the FMO phonon spectral density on excitonic dynamics over the simulated picosecond timescale~\cite{LorenzoniPRL2024} (see the SM for details). This suggests that phonon-mediated transitions between the exciton states occur not only via resonant vibrational modes with frequencies close to the electronic energy gap $\Delta_{36}$, but also through interactions with the entire vibrational environment~\cite{Caycedo2022}. Thus, vibronic interactions in the FMO complex go beyond the weak coupling regime and long-lived excitonic coherences are supported by the entire highly-structured phonon spectral density, instead of a few resonant vibrational modes, revealing the limitations of conventional theoretical approaches~\cite{KolliJCP2012,TiwariPNAS2013,ChinNP2013}.

\begin{figure}
\includegraphics[width=0.38\textwidth]{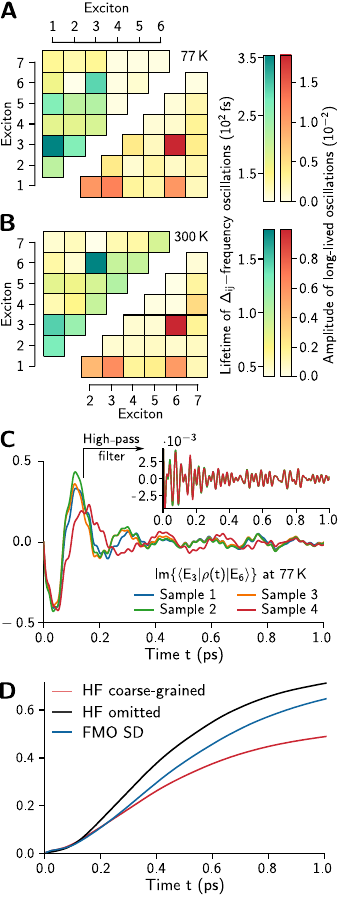}
\caption{\textsf{\textbf{Long-lived excitonic coherences and their robustness against static disorder.} The lifetimes of the large-amplitude oscillations with frequencies $\Delta_{ij}$ and the amplitudes of long-lived multi-frequency oscillations in the excitonic coherence dynamics $\langle E_i|\rho(t)|E_j\rangle$ are shown at (A) $77\,{\rm K}$ and (B) $300\,{\rm K}$. (C) The excitonic coherence dynamics $\langle E_3|\rho(t)|E_6\rangle$ between $\ket{E_3}$ and $\ket{E_6}$ at $77\,{\rm K}$, computed with four randomly generated sets of electronic parameters based on a Gaussian static disorder model, are shown. The coherence dynamics exhibit essentially identical high-frequency oscillations,  as shown in the inset, obtained via a high-pass filter. (D) For the initial state $\ket{E_7}$, the population dynamics of $\ket{E_1}$ are present under three different phonon spectral densities (SDs): the microscopic FMO SD, and two model SDs in which high-frequency (HF) intra-pigment modes beyond $700\,{\rm cm}^{-1}$ are either omitted or coarse-grained. See the SM for more details.}}
\label{Fig4}
\end{figure}

For all possible exciton pairs $(j,k)$, Fig.~4A summarizes the lifetimes of the large-amplitude oscillations of excitonic coherences with frequencies close to $\Delta_{jk}$, along with the amplitudes of high-frequency long-lived oscillations. We found that the coherence between exciton states exhibits the long-lived multi-frequency characteristics when at least one of the states is either the lowest-energy exciton $\ket{E_1}$ or the meta-stable state $\ket{E_3}$ whose population transfer to other exciton states remains small within $1\,{\rm ps}$ (see the SM). This observation was further confirmed by modifying electronic parameters, varying the number of meta-stable states (see the SM). Notably, we found that while the frequencies of the large-amplitude oscillations depend on the randomness of electronic parameters induced by static disorder, the high-frequency oscillations remain nearly unchanged, as shown in Fig.~4B. These results indicate that long-lived excitonic coherences supported by multi-mode vibrational environments of PPCs do not require fine-tuning of electronic energy-level structures, making the observed quantum effects relevant in biological conditions.

We note that intra-pigment modes with frequencies beyond the energy range of excitonic splittings ($\Delta_{ij}<700\,{\rm cm}^{-1}$) can significantly influence EET, even though high-frequency excitonic coherences exhibit small amplitudes. This is due to quantum effects in photosynthetic EET arising from superpositions between electronic-vibrational states of photosynthetic excitons and intra-pigment vibrational modes, which are only partially revealed by monitoring excitonic coherence dynamics. Fig.~4D presents the population dynamics of the lowest-energy exciton $\ket{E_1}$ when the highest-energy exciton $\ket{E_7}$ is initially populated. In addition to the full FMO phonon spectral density, we examine two model spectral densities in which protein and intra-pigment modes up to $700\,{\rm cm}^{-1}$ are treated accurately, while all other high-frequency intra-pigment modes are either omitted or coarse-grained with an increased vibrational damping rate $(50\,{\rm fs})^{-1}$. The population dynamics show notable differences depending on how these high-frequency modes are handled.

\textbf{2D electronic spectroscopy}

\begin{figure*}
\includegraphics[width=0.66\textwidth]{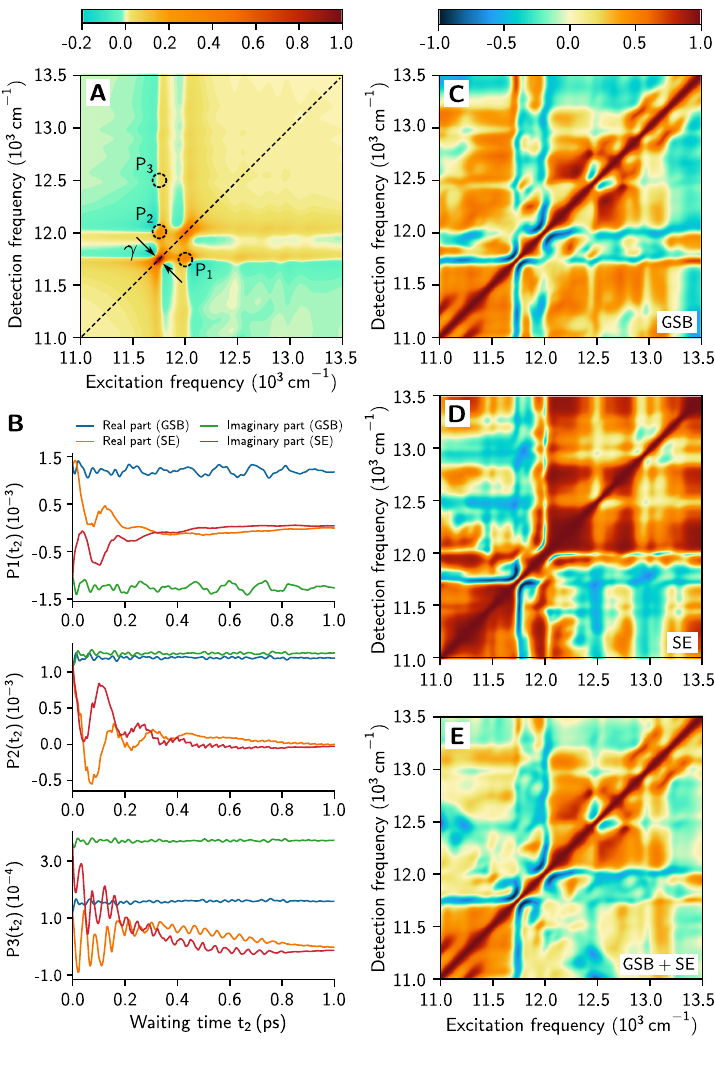}
\caption{\textsf{\textbf{Numerically exact 2D electronic spectra.} (A) The rephasing spectra of a heterodimer at $t_2=0$ are shown with $\gamma$ denoting the anti-diagonal width of a diagonal peak. (B) For the three peak positions marked in (A), the peak dynamics as functions of the waiting time $t_2$ are displayed. The correlation maps of (C) ground-state bleaching (GSB), (D) stimulated emission (SE), and (E) the total 2D signals (GSB+SE) are shown, computed based on the real parts of the long-lived oscillatory signals over a finite time window of $300\,{\rm fs}\le t_2 \le 1\,{\rm ps}$. In simulations, the microscopic FMO phonon spectral density at $77\,{\rm K}$ was considered (see Fig.~2). See the SM for more details.}}
\label{Fig5}
\end{figure*}

So far, we have theoretically demonstrated that excitonic coherences in the FMO complex are long-lived in realistic vibrational environments. To clarify that our results do not contradict the experimental evidence used to claim the absence of quantum effects in the FMO complex, we revisit the measures discussed in Ref.~\cite{DuanPNAS2017}. 

In 2D electronic spectroscopy~\cite{JonasARPC2003,BrixnerJCP2004}, the third-order molecular polarization induced by a sequence of laser pulses is measured as a function of time delay $t_2$ between pump and probe pulses. By using a pair of pump pulses with controlled time delay, 2D electronic spectra are measured as a function of excitation and detection frequencies ($\omega_1$,$\omega_3$) for each waiting time $t_2$. In particular, for the rephasing 2D electronic spectra at $t_2 = 0$, the width of diagonal peaks in the anti-diagonal direction allows an estimation of the lifetime of optical coherence (i.e between the electronic ground and excited states) at the single-molecule level, thereby separating the homogeneous broadening from the inhomogenous broadening caused by static disorder. In Ref.~\cite{DuanPNAS2017}, the experimentally measured width of around $60\,{\rm fs}$ was considered one piece of evidence supporting the absence of quantum effects in the FMO complex at room temperature, supplied by an approximate theoretical approach employing the coarse-grained spectral density shown in Fig.~2. However, the lifetime of optical coherences does not directly determine that of excitonic coherences, as the relationship between them depends strongly on the nature of the electron-phonon interaction.

To clarify this issue, we simulated the rephasing 2D spectra of a dimeric PPC model at $77\,{\rm K}$, computed non-perturbatively using the FMO phonon spectral density (see the SM for details). The total 2D spectra are expressed as the sum of the stimulated emission (SE) and ground-state bleaching (GSB) signals, which provide information about the molecular dynamics within the electronic excited-state and ground-state manifolds, respectively, during the waiting time $t_2$~\cite{BrixnerJCP2004,ChoBook2009,Lim2019}. While the SE and GSB signals can be computed separately, only the total signal can be measured in experiments. Therefore, by comparing the lifetime of oscillatory SE signals, originating from excited-state coherences, with the anti-diagonal width of a diagonal peak in rephasing 2D spectra at $t_2 = 0$, one can verify whether the experimental measure considered in Ref.~\cite{DuanPNAS2017} is reliable. Fig.~5A displays the numerically exact rephasing spectra at $t_2 = 0$, where the anti-diagonal width is approximately $200\,{\rm fs}$. As shown in Fig.~5B, the oscillatory features in the SE signals last up to $1\,{\rm ps}$, demonstrating that the lifetime of excited-state coherences appearing in 2D spectra is not limited by the anti-diagonal width.

As another evidence, Ref.~\cite{DuanPNAS2017} examined the correlation between oscillatory 2D signals for each pair of spectral positions ($\omega_1$,$\omega_3$) and ($\omega_3$,$\omega_1$), which are symmetric with respect to the diagonal line $\omega_1=\omega_3$. The relative phase between these oscillations can be quantified using the Pearson correlation coefficient, where positive and negative values indicate in-phase and out-of-phase relationships, respectively. Based on approximate theoretical results~\cite{ButkusCPL2012}, it was hypothesized that negative correlations serve as signatures of ground-state coherences, since excited-state coherences were expected to produce only positive correlations. However, this hypothesis does not hold for realistic PPC models. Figs.~5C and D present the correlation maps of the SE and GSB signals, respectively. Both maps display regions of positive and negative correlations depending on spectral positions, indicating that negative correlations do not necessarily imply the oscillatory GSB signals originating from ground-state coherences. As shown in Fig.~5E, the correlation map of the total 2D spectra exhibits negative correlations at cross peaks and positive correlations at diagonal peaks. This pattern closely resembles the experimentally measured correlation map of the FMO complex reported in Ref.~\cite{DuanPNAS2017}. Since long-lived excited-state coherences can be present at the negatively correlated cross peaks (see Figs.~5B and E), the negative correlations cannot be taken as evidence for the absence of excited-state coherences.

\textbf{Conclusions}

Our results demonstrate that quantum effects in the FMO complex persist up to $1\,{\rm ps}$ and beyond, which is the time scale of photosynthetic excitation energy transfer. These conclusions are based on electronic dynamics rigorously computed using a microscopic model constructed from first-principles methods~\cite{CokerJPCL2016,RheePCCP2018,KleinekathoferPR2023} and validated with experimental spectroscopic data, such as linear absorption, linear and circular dichroism~\cite{VultoJPCB1998,RengerBJ2006}, and fluorescence line narrowing~\cite{RatsepJL2007}. More broadly, our findings suggest that similar quantum effects may exist in other pigment-protein complexes, as highly-structured phonon spectral densities are a ubiquitous feature arising from the vibrational normal modes of pigments~\cite{ZazubovichJPCB2001,OgilvieJPCL2018,HarelJPCL2018,ColliniCP2019}.

Our 2D simulations indicate that verifying or falsifying quantum effects in PPCs cannot be reliably achieved using nonlinear spectroscopic data analyzed with approximate theoretical methods. While various nonlinear spectroscopic techniques have been designed to isolate excited-state coherences, such as through controlled laser pulse polarizations~\cite{ThyrhaugNC2018} or colors~\cite{SenlikJPCL2015}, their performance in eliminating ground-state coherences remains insufficient for investigating quantum effects in PPCs~\cite{Lim2019}. Therefore, a quantitative, rather than qualitative, analysis of nonlinear spectroscopic data using microscopic PPC models is essential for experimentally studying quantum effects in PPCs, an approach that has not yet been explored in current state-of-the-art research. Alternatively, the development of new experimental techniques may be required, as most nonlinear spectroscopic techniques measure the optical responses of molecular ensembles and may underestimate quantum effects in photosynthetic EET occurring at the single-molecule level.

Ultimately, this approach will provide a new framework for interpreting experimental data on photosynthetic pigment-protein complexes, where electronic dynamics occur in non-perturbative regimes. This may reveal how photosynthetic excitation energy transfer occurs in real biological environments and how quantum effects are sustained and utilized under ambient conditions.

\textbf{Acknowledgments}: We thank Thomas Renger, Chang Woo Kim, and Young Min Rhee for helpful discussions and for providing the FMO parameters.

\textbf{Funding}: This work was support by the ERC Synergy grant HyperQ (Grant No. 856432), the BMBF project PhoQuant (grant no. 13N16110), the state of Baden-Württemberg through bwHPC and the German Research Foundation (DFG) through grant no INST 40/575-1 FUGG (JUSTUS 2 cluster), and the Next Generation EU via the NQSTI-Spoke1-BaC project QBETTER (contract n. PE 0000023-QBETTER).
 
\textbf{Competing Interests}:  The authors declare they have no competing interest.

\textbf{Data and Materials Availability}: Details about the model and numerical tools are provided in the supplementary materials. Simulated data that supports the finding are available upon request.

\newpage
\begin{widetext}

\begin{center}
\textbf{Supplementary materials}
\end{center}

\section{Materials and Methods}

\subsection{Microscopic model of the FMO photosynthetic complex}

\begin{table}
\caption{Electronic Hamiltonian of the FMO complex, estimated in Ref.~\cite{SM_RengerBJ2006}.}
\label{Table_electronic_parameters}
\begin{ruledtabular}
\begin{tabular}{rrrrrrrr}
$\sf{\langle n|H_e|m\rangle (cm^{-1})}$ & 1 & 2 & 3 & 4 & 5 & 6 & 7\\
\hline\hline
1& 12410 & -87.7 & 5.5 & -5.9 & 6.7 & -13.7 & -9.9\\
\hline
2&       & 12530 & 30.8 & 8.2 & 0.7 & 11.8 & 4.3\\
\hline
3&       &       & 12210 & -53.5 & -2.2 & -9.6 & 6.0\\
\hline
4&       &       &       & 12320 & -70.7 & -17.0 & -63.3\\
\hline
5&       &       &       &       & 12480 & 81.1 & -1.3\\
\hline
6&       &       &       &       &       & 12630 & 39.7\\
\hline
7&       &       &       &       &       &       & 12440\\
\end{tabular}
\end{ruledtabular}
\end{table}

\begin{table}
\caption{\textsf{The vibrational frequencies $\omega_k$ and the Huang-Rhys factors $s_k$ of the 62 intra-pigment modes of the FMO complex~\cite{SM_RatsepJL2007}.}}
\label{Table_intrapigment}
\begin{ruledtabular}
\begin{tabular}{llllllllllllll}
$k$ & 1 & 2 & 3 & 4 & 5 & 6 & 7 & 8 & 9 & 10\\
\hline
$\omega_k\,[{\rm cm}^{-1}]$ & 46 & 68 & 117 & 167 & 180 & 191 & 202 & 243 & 263 & 284 \\
$s_k$ & 0.011 & 0.011 & 0.009 & 0.009 & 0.010 & 0.011 & 0.011 & 0.012 & 0.003 & 0.008 \\
\hline\hline
$k$ & 11 & 12 & 13 & 14 & 15 & 16 & 17 & 18 & 19 & 20 \\
\hline
$\omega_k\,[{\rm cm}^{-1}]$ & 291 & 327 & 366 & 385 & 404 & 423 & 440 & 481 & 541 & 568 \\
$s_k$ & 0.008 & 0.003 & 0.006 & 0.002 & 0.002 & 0.002 & 0.001 & 0.002 & 0.004 & 0.007\\
\hline\hline
$k$ & 21 & 22 & 23 & 24 & 25 & 26 & 27 & 28 & 29 & 30 \\
\hline
$\omega_k\,[{\rm cm}^{-1}]$ & 582 & 597 & 630 & 638 & 665 & 684 & 713 & 726 & 731 & 750 \\
$s_k$ & 0.004 & 0.004 & 0.003 & 0.006 & 0.004 & 0.003 & 0.007 & 0.010 & 0.005 & 0.004 \\
\hline\hline
$k$ & 31 & 32 & 33 & 34 & 35 & 36 & 37 & 38 & 39 & 40\\
\hline
$\omega_k\,[{\rm cm}^{-1}]$ & 761 & 770 & 795 & 821 & 856 & 891 & 900 & 924 & 929 & 946 \\
$s_k$ & 0.009 & 0.018 & 0.007 & 0.006 & 0.007 & 0.003 & 0.004 & 0.001 & 0.001 & 0.002\\
\hline\hline
$k$ & 41 & 42 & 43 & 44 & 45 & 46 & 47 & 48 & 49 & 50 \\
\hline
$\omega_k\,[{\rm cm}^{-1}]$ & 966 & 984 & 1004 & 1037 & 1058 & 1094 & 1104 & 1123 & 1130 & 1162 \\
$s_k$ & 0.002 & 0.003 & 0.001 & 0.002 & 0.002 & 0.001 & 0.001 & 0.003 & 0.003 & 0.009 \\
\hline\hline
$k$ & 51 & 52 & 53 & 54 & 55 & 56 & 57 & 58 & 59 & 60 \\
\hline
$\omega_k\,[{\rm cm}^{-1}]$ & 1175 & 1181 & 1201 & 1220 & 1283 & 1292 & 1348 & 1367 & 1386 & 1431 \\
$s_k$ & 0.007 & 0.010 & 0.003 & 0.005 & 0.002 & 0.004 & 0.007 & 0.002 & 0.004 & 0.002 \\
\hline\hline
$k$ & 61 & 62 \\
\hline
$\omega_k\,[{\rm cm}^{-1}]$ & 1503 & 1545 \\
$s_k$ & 0.003 & 0.003 \\
\end{tabular}
\end{ruledtabular}
\end{table}

\begin{table}
\caption{\textsf{Transition dipole moments of the seven pigments in the FMO complex. The parameters are given in arbitrary units, as only relative values are relevant in this work.}}
\label{Table_dipole_moments}
\begin{ruledtabular}
\begin{tabular}{rrrrrrrr}
$\bm{\mu}_n$ & 1 & 2 & 3 & 4 & 5 & 6 & 7 \\
\hline\hline
$x$-component & -0.741 & -0.857 & -0.197 & -0.799 & -0.740 & -0.135 & -0.495\\
\hline
$y$-component &  -0.561 & 0.504 & 0.957 & -0.534 & 0.656 & -0.879 & -0.708\\
\hline
$z$-component & -0.370 & -0.107 & -0.211 & -0.277 & 0.164 & 0.457 & -0.503\\
\end{tabular}
\end{ruledtabular}
\end{table}

The Hamiltonian of the FMO complex is modeled by $H_e + H_v + H_{e-v}$, where $H_e$ is the electronic Hamiltonian, $H_v$ is the vibrational Hamiltonian, and $H_{e-v}$ describes the interaction Hamiltonian between electronic states and vibrational modes. Each pigment is modeled as a two-level system, as the energy gap between its first and second excited states is sufficiently large to allow selective excitation of the first excited state using a laser pulse. We consider the electronic ground and singly-excited state manifolds, as these states govern photosynthetic excitation energy transfer, and the linear and nonlinear optical spectra.

The electronic Hamiltonian is given by
\begin{equation}
    H_e=\sum_{n=1}^{7}\epsilon_n\ket{\epsilon_n}\bra{\epsilon_n}+\sum_{n\neq m}^{7}V_{nm}\ket{\epsilon_n}\bra{\epsilon_m},
\end{equation}
where $\ket{\epsilon_n}$ represents the local electronic excitation at site $n$ with site energy $\epsilon_n$, while $V_{nm}$ denotes the electronic coupling between sites $n$ and $m$. In the main text, we consider the electronic parameters of the FMO complex estimated in Ref.~\cite{SM_RengerBJ2006}, summarized in Table~\ref{Table_electronic_parameters}.

The vibrational modes are modeled as independent quantum harmonic oscillators, with the vibrational Hamiltonian
\begin{equation}
    H_v=\sum_{n=1}^{7}\sum_{\xi}\omega_\xi b_{n,\xi}^{\dagger}b_{n,\xi},
\end{equation}
where $b_{n,\xi}^{\dagger}$ and $b_{n,\xi}$ are the creation and annihilation operators of the vibrational mode with frequency $\omega_\xi$, locally coupled to site $n$. The interaction Hamiltonian is expressed as
\begin{equation}
H_{e-v}=\sum_{n=1}^{7}\ket{\epsilon_n}\bra{\epsilon_n}\sum_{\xi}\omega_\xi\sqrt{s_\xi}(b_{n,\xi}+b_{n,\xi}^{\dagger}),
\end{equation}
where the vibronic coupling strength is quantified by the Huang-Rhys factor $s_\xi$.

In this work, we present excitonic population and coherence dynamics, obtained by non-perturbatively simulating ${\rho(t)={\rm Tr}_v [e^{-iHt}\rho(0)\rho_{v}(T)e^{iHt}]}$, where $\rho(t)$ is the reduced electronic density matrix at time $t$, ${\rm Tr}_v$ denotes the partial trace over vibrational degrees of freedom, $\rho(0)$ is the initial electronic state, and $\rho_{v}(T)=\exp(-H_v/k_B T)/{\rm Tr}[\exp(-H_v/k_B T)]$ is the thermal state of the vibrational environment at temperature $T$. In the excitonic coherence simulations, shown in Fig.~3 and 4C of the main text, the initial electronic state is modeled as $\rho(0) = |\psi\rangle\langle \psi|$ with $|\psi\rangle \propto \sum_{n=1}^{7} (\bm{\mu}_n\cdot \bm{e}_{\rm Laser})|\epsilon_n\rangle$, where $\bm{\mu}_n$ denotes the transition dipole moment of site $n$ in the FMO complex, summarized in Table~\ref{Table_dipole_moments}, and $\bm{e}_{\rm Laser}=(0.25, 0.85, 0.0)$ represents the laser pulse polarization, chosen to populate all seven exciton states $\ket{E_k}$ with comparable magnitudes. In the exciton population simulations shown in Fig.~4D of the main text, the initial state is assumed to be the highest-energy exciton state, $\ket{\psi}=\ket{E_7}$.

The influence of the vibrational environment on the reduced electronic density matrix $\rho(t)$ is fully characterized by the phonon spectral density defined as $J(\omega)=\sum_{\xi}\omega_{\xi}^{2}s_\xi\delta(\omega-\omega_\xi)$. The experimentally estimated phonon spectral density of the FMO complex~\cite{SM_RatsepJL2007} is given by $J(\omega) = J_{\rm AR}(\omega)+J_{L}(\omega)$, where $J_{\rm AR}(\omega)$ represents the Adolphs-Renger (AR) spectral density describing protein modes
\begin{equation}
	J_{\rm AR}(\omega)=\frac{S_0}{S_1+S_2}\sum_{i=1}^{2}\frac{S_i}{7! 2 \Omega_i^4}\omega^5 e^{-(\omega/\Omega_i)^{1/2}},
	\label{eq:B777_SD}
\end{equation}
with $S_0=0.29$, $S_1=0.8$, $S_2=0.5$, $\Omega_1=0.069\,{\rm meV}$ and $\Omega_2=0.24\,{\rm meV}$, while $J_{L}(\omega)$ denotes the sum of 62 Lorentzian spectral densities corresponding to intra-pigment modes
\begin{equation}
    J_{L}(\omega) = \sum_{k=1}^{62} \frac{4 \omega_k s_k \gamma_k (\omega_{k}^2+\gamma_{k}^2)\omega}{\pi((\omega+\omega_k)^{2}+\gamma_{k}^2)((\omega-\omega_k)^{2}+\gamma_{k}^2)}.
\end{equation}
Each Lorentzian spectral density is characterized by the vibrational frequency $\omega_k$ and Huang-Rhys factor $s_k$ of the $k$-th intra-pigment mode, as summarized in Table~\ref{Table_intrapigment}. The vibrational damping rates of the intra-pigment modes are taken to be $\gamma_k = (1\,{\rm ps})^{-1}\approx 5\,{\rm cm}^{-1}$. The reorganization energy of the intra-pigment modes is given by $\int_{0}^{\infty}d\omega J_{L}(\omega)/\omega = \sum_{k=1}^{62}\omega_k s_k$, which is independent of the vibrational damping rates $\gamma_k$. The total reorganization energy of the FMO phonon spectral density $J(\omega)$ is $\int_{0}^{\infty}d\omega J(\omega)/\omega\approx 232\,{\rm cm}^{-1}$.

\subsection{Model phonon spectral densities with approximated high-frequency intra-pigment modes}

\begin{figure}[t]
\includegraphics[width=0.66\textwidth]{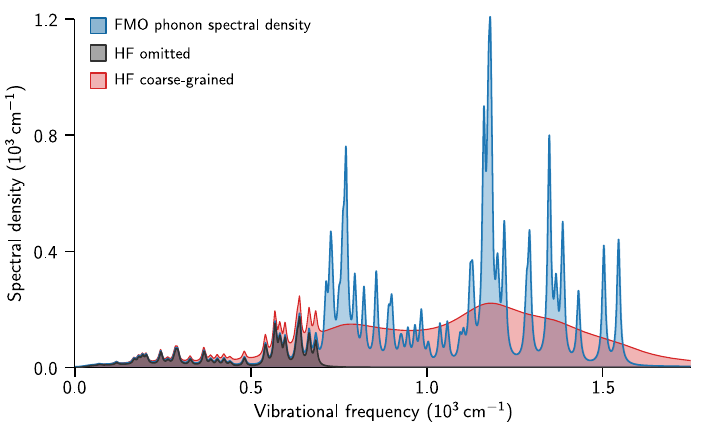}
\caption{\textbf{Approximated high-frequency intra-pigment modes}. The model spectral densities, with high-frequency intra-pigment modes (vibrational frequencies $\omega_k\ge 700\,{\rm cm}^{-1}$) either omitted (black) or coarse-grained (red), are shown. The experimentally estimated FMO phonon spectral density is shown in blue.}
\label{FigS1}
\end{figure}

In Fig.~4D of the main text, we consider two model spectral densities where the first 26 intra-pigment modes with $\omega_k<700\,{\rm cm}^{-1}$ are considered accurately using the parameters in Table~\ref{Table_intrapigment} with $\gamma_k=(1\,{\rm ps})^{-1}$, while the other high-frequency intra-pigment modes are either omitted ($s_k = 0$ for $k>26$) or coarse-grained ($\gamma_k = (50\,{\rm fs})^{-1}$ for $k>26$), as shown in black and red, respectively, in Fig.~\ref{FigS1}. In both cases, the AR spectral density $J_{\rm AR}(\omega)$ is considered. The total reorganization energy of the coarse-grained case is identical to that of the FMO phonon spectral density.

\subsection{Coarse-grained phonon spectral density}

In Figs.~2 and 3 of the main text, we consider the coarse-grained phonon spectral density from Ref.~\cite{SM_DuanPNAS2017}
\begin{equation}
    J_{\rm CG}(\omega)=\frac{2}{\pi}\gamma \omega e^{-\omega/\omega_c}+\frac{4}{\pi^2}S\Omega^3 \frac{\omega\Gamma}{(\Omega^2-\omega^2)^2 + \omega^2 \Gamma^2},
\end{equation}
where $\gamma=0.7$, $\omega_c=350\,{\rm cm}^{-1}$, $S=0.12$, $\Omega=900\,{\rm cm}^{-1}$, and $\Gamma = 700\,{\rm cm}^{-1}$, resulting in a reorganization energy of approximately $224\,{\rm cm}^{-1}$. We note that these parameters were confirmed by the first author of Ref.~\cite{SM_DuanPNAS2017} through private communication.

\subsection{Refined electronic parameters of the FMO complex}

\begin{figure*}
\includegraphics[width=0.66\textwidth]{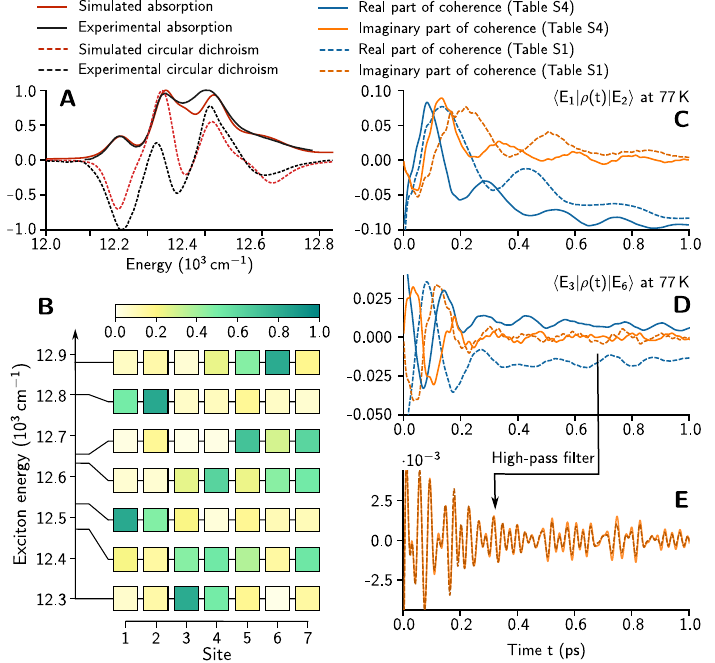}
\caption{\textsf{\textbf{Refined electronic parameters of the FMO complex and excitonic coherence dynamics.} (A) Numerically exact absorption and circular dichroism spectra of the FMO complex at $77\,{\rm K}$, computed using Table~\ref{Table_electronic_parameters_refined} and the experimentally estimated phonon spectral density of the FMO complex, are shown in red. These computed spectra closely match the experimental data~\cite{SM_MelkozernovPhotosynthRes1998} shown in black. (B) Population distributions of the seven exciton states $\ket{E_k}$ of the FMO complex in the site basis, computed using the refined electronic parameters in Table~\ref{Table_electronic_parameters_refined}. (C) The excitonic coherence dynamics at $77\,{\rm K}$ between $\ket{E_1}$ and $\ket{E_2}$ and (D) those between $\ket{E_3}$ and $\ket{E_6}$, simulated using the refined electronic parameters, are shown as solid lines. For comparison with Fig.~3 of the main text, the excitonic coherence dynamics computed using the electronic parameters estimated in Ref.~\cite{SM_RengerBJ2006} are shown in dashed lines. (E) The high-frequency long-lived components of the excitonic coherences between $\ket{E_3}$ and $\ket{E_6}$, extracted via a high-pass filter, are nearly identical for both sets of electronic parameters.}}
\label{FigS2}
\end{figure*}

\begin{table}
\caption{\textsf{Refined electronic Hamiltonian of the FMO complex, obtained by fitting numerically exact absorption and circular dichroism spectra to experimental data at $T=77\,{\rm K}$ (see Fig.~\ref{FigS2}A).}}
\label{Table_electronic_parameters_refined}
\begin{ruledtabular}
\begin{tabular}{rrrrrrrr}
$\sf{\langle n|H_e|m\rangle (cm^{-1})}$ & 1 & 2 & 3 & 4 & 5 & 6 & 7\\
\hline\hline
1& 12598 & -120.0 & 8.2 & -8.5 & 10.0 & -13.1 & -16.1\\
\hline
2&       & 12719 & 46.8 & 16.3 & 1.7 & 12.9 & 9.7\\
\hline
3&       &       & 12372 & -99.8 & -13.0 & -2.8 & 6.1\\
\hline
4&       &       &       & 12515 & -93.0 & -38.1 & -80.0\\
\hline
5&       &       &       &       & 12664 & 102.5 & -31.9\\
\hline
6&       &       &       &       &       & 12794 & 53.8\\
\hline
7&       &       &       &       &       &       & 12593\\
\end{tabular}
\end{ruledtabular}
\end{table}

The electronic parameters of the FMO complex have been estimated using first-principles methods and fine-tuned based on a comparison of theoretically computed linear optical spectra with experimental data, such as linear absorption and circular dichroism~\cite{SM_RengerBJ2006}. However, the linear optical spectra have been computed using approximate methods, thereby a refinement of electronic parameters based on non-perturbative simulations is needed.

Fig.~\ref{FigS2}A shows the numerically exact absorption and circular dichorism spectra at $T=77\,{\rm K}$, computed based on a refined set of electronic parameters summarized in Table~\ref{Table_electronic_parameters_refined}. Here, the electronic parameters were optimized to quantitatively reproduce the experimental data.

For the refined set of electronic parameters (i.e. Table~\ref{Table_electronic_parameters_refined}), Fig.~\ref{FigS2}B summarizes the properties of exciton states (i.e. the electronic eigenstates of $H_e$), which are qualitatively similar to those shown in Fig.~1C of the main text, obtained using the electronic parameters estimated in Ref.~\cite{SM_RengerBJ2006} (i.e. Table~\ref{Table_electronic_parameters}).

In Fig.\ref{FigS2}C (D), the excitonic coherence dynamics between $\ket{E_1}$ and $\ket{E_2}$ ($\ket{E_3}$ and $\ket{E_6}$) at $T=77\,{\rm K}$, non-perturbatively computed based on the refined set of electronic parameters, are shown in solid lines. For comparison with Fig.~3 of the main text, the excitonic coherence dynamics computed using the electronic parameters estimated in Ref.~\cite{SM_RengerBJ2006} are shown as dashed lines. For both sets of electronic parameters, the excitonic coherences exhibit large-amplitude oscillations with a frequency close to the energy-gap $\Delta_{ij} = |E_i - E_j|$ between exciton states, along with high-frequency long-lived oscillations of relatively small amplitudes. Notably, the high-frequency long-lived components of the excitonic coherence between $\ket{E_3}$ and $\ket{E_6}$ are nearly identical for both parameter sets. This is illustrated in Fig.~\ref{FigS2}E, where a high-frequency filter was applied to remove the large-amplitude oscillations associated with the lower-frequency $\Delta_{36}$. In addition to Fig.~4C of the main text, Fig.~\ref{FigS2}E further demonstrates that the long-lived excitonic coherences of the FMO complex are robust against variations in electronic parameters.

\subsection{Non-perturbative simulation method}

\begin{figure}[t]
\includegraphics[width=0.66\textwidth]{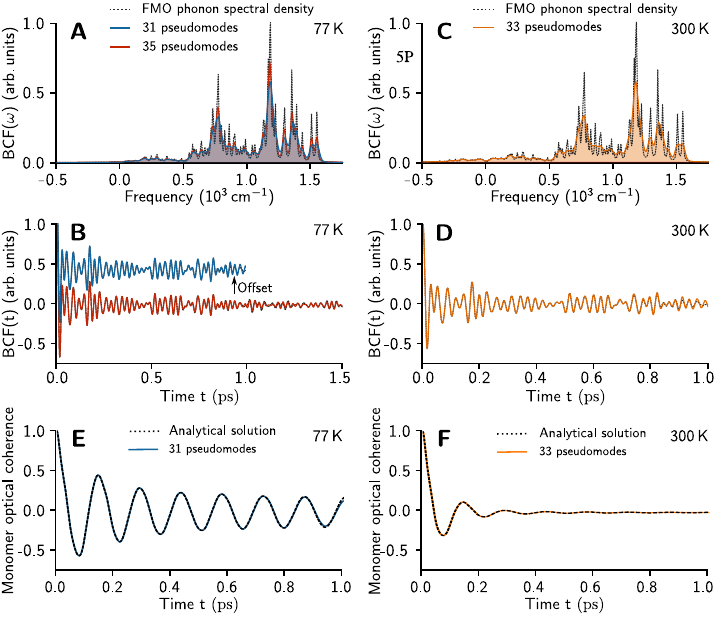}
\caption{\textsf{\textbf{Systematic coarse-graining of the environment.} (A) The frequency spectra of the bath correlation functions (BCFs) of the effective environments modeled using 31 and 35 pseudomodes are shown in blue and red, respectively, compared with the FMO phonon spectral density (SD) shown as a black dashed line. (B) The effective environments quantitatively reproduce the BCF of the FMO SD at $77\,{\rm K}$ up to $1.0\,{\rm ps}$ and $1.5\,{\rm ps}$, respectively. (C) The frequency spectrum of the BCF of the effective environment modeled using 33 pseudomodes, including two negative-frequency modes, is shown in orange. (D) This effective environment can quantitatively reproduce the BCF of the FMO SD at $300\,{\rm K}$ up to $1.0\,{\rm ps}$. Numerically exact optical coherence dynamics of a monomer computed by DAMPF, based on the pseudomode parameters, are well matched with the analytical solutions of the monomer dynamics at (E) $77\,{\rm K}$ and (F) $300\,{\rm K}$.}}
\label{FigS3}
\end{figure}

Non-perturbatively simulating the dynamics of open quantum systems at finite temperatures is a formidable challenge in the presence of highly-structured spectral densities, as those typical of PPCs. In this work, we tackle this scenario by employing the Dissipation-Assisted Matrix Product Factorization (DAMPF) method~\cite{SM_SomozaPRL2019}, a non-perturbative, numerically exact approach that has been successfully applied to the simulation of vibronic systems~\cite{SM_SomozaCommunPhys2023}. We further enhance it by incorporating the systematic coarse-graining of environments introduced in Ref.~\cite{SM_LorenzoniPRL2024} and the concept of thermalized spectral density from Ref.~\cite{SM_TamascelliPRL2019}.

In particular, since the electronic degrees of freedom linearly couple to a collection of quantum harmonic oscillators representing the vibrational environment of PPCs, and the thermal state $\rho_{v}(T)=\exp(-H_v/k_B T)/{\rm Tr}[\exp(-H_v/k_B T)]$ is taken as the initial state of the environment, it follows from Ref.~\cite{SM_TamascelliPRL2018} that the original environmental modes can be replaced by a finite number of damped quantum harmonic oscillators (pseudomodes), as both yield an identical reduced electronic density matrix $\rho(t)$, provided that their two-time bath correlation functions (BCFs) are well matched~\cite{SM_SomozaPRL2019}. Since energy transfer in the FMO complex occurs on a characteristic timescale of $\tau\simeq\SI{1}{ps}$, we perform numerically exact simulations of excitonic population and coherence dynamics of the FMO complex up to $1\,{\rm ps}$. This allows us to apply the systematic environmental coarse-graining technique of Ref.~\cite{SM_LorenzoniPRL2024}, enabling the use of a simplified, effective environment that accurately reproduces the BCF of the FMO phonon spectral density, defined as ${\rm BCF}(t)=\int_{0}^{\infty}d\omega J(\omega)(\coth(\omega/2k_B T)\cos(\omega t)-i\sin(\omega t))$, over the finite time window $0\le t\le 1\,{\rm ps}$. In Fig.~\ref{FigS3}A, we show the frequency spectrum of the BCF at $T=\SI{77}{K}$ of the effective environment modeled using 31 pseudomodes for DAMPF simulations. Fig.~\ref{FigS3}B demonstrates that the BCFs of the FMO phonon spectral density and the effective pseudomode environment are well matched up to $\tau\simeq\SI{1}{ps}$. This ensures that the reduced electronic density matrix $\rho(t)$ can be computed accurately within the finite time window $0\le t\le 1\,{\rm ps}$ using DAMPF.

For non-perturbative simulations of two-dimensional rephasing spectra of a dimeric PPC model at $T=77\,{\rm K}$, we considered an extended time window $0\le t\le 1.5\,{\rm ps}$, leading to an effective environment modeled using 35 pseudomodes, as shown in Fig.~\ref{FigS3}A, which accurately reproduces the BCF of the FMO phonon spectral density up to $1.5\,{\rm ps}$, as shown in Fig.~\ref{FigS3}B. In 2D simulations, light-matter interaction was treated perturbatively to compute the ground-state bleaching and stimulated emission signals separately. This assumption is valid since the laser intensity is reduced in 2D experiments until the rephasing spectra show convergence, ensuring that experimental data are dominated by third-order optical responses with negligible higher-order contributions. We considered a heterodimer with a site energy difference of $\epsilon_1-\epsilon_2=200\,{\rm cm}^{-1}$ and an inter-site electronic coupling of $V_{12}=100\,{\rm cm}^{-1}$. The transition dipole moments of the two pigments were assumed to be orthogonal, similar to sites 1 and 2 of the FMO complex, and an orientational average with respect to the fixed laser polarization was taken into account in simulations. As in the case of the FMO complex, we assumed that each site of the dimer is coupled to a local vibrational environment modeled by the FMO phonon spectral density.

At room temperature $T=300\,{\rm K}$, DAMPF simulations become significantly more challenging due to the increased local dimensions of the pseudomodes caused by higher thermal populations. To achieve the same level of accuracy as in the $T=77\,{\rm K}$ simulations, we improved the DAMPF method by incorporating the concept of thermalized spectral density introduced in Ref.~\cite{SM_TamascelliPRL2019}. This approach effectively describes a finite-temperature harmonic environment using a zero-temperature harmonic environment while accurately capturing the influence of finite temperature on the reduced open system dynamics, at the cost of enhanced couplings and the introduction of negative-frequency harmonic modes. When this idea is directly applied to DAMPF, one can consider zero-temperature pseudomodes with either positive or negative frequencies. However, at lower temperatures, correlations between pseudomodes tend to increase, which can lead to higher bond dimensions and increased computational costs. Thus, there exists an optimal temperature range where the thermal populations of the pseudomodes are lower than at room temperature, while the correlations between pseudomodes are lower than at zero temperature. In this work, we considered a partial thermal spectral density by independently optimizing the temperatures of the pseudomodes, along with their frequencies, decay rates and the Huang-Rhys factors when fitting the BCF of the FMO phonon spectral density at $T=300\,{\rm K}$. As in the case of the full thermalized spectral density~\cite{SM_TamascelliPRL2019}, the partial thermal spectral density increases the couplings between electronic states and pseudomodes and requires the use of negative frequencies for some pseudomodes, but both the local and bond dimensions are significantly reduced, boosting numerical efficiency. In Fig.~\ref{FigS3}C, 
we show the frequency spectrum of the BCF at $T=300\,{\rm K}$ of the effective environment modeled using 33 pseudomodes for DAMPF simulations, including two modes with negative frequencies. Fig.~\ref{FigS3}D demonstrates that the BCFs are well matched up to $\tau\simeq\SI{1}{ps}$, ensuring that the room-temperature EET dynamics of the FMO complex can be simulated with high accuracy using DAMPF.

To test the accuracy of the pseudomode parameters prepared for $77\,{\rm K}$ and $300\,{\rm K}$ DAMPF simulations of the FMO complex, we compared analytical solutions of the optical coherence dynamics of a monomer, where a single site couples to the FMO phonon spectral density, with numerical results obtained by DAMPF. As shown in Figs.~\ref{FigS3}E and F, the numerically exact monomer dynamics obtained by DAMPF for $77\,{\rm K}$ and $300\,{\rm K}$ are well matched with the analytical solutions. We note that our in-house DAMPF code has been benchmarked against other in-house codes implementing independent non-perturbative methods~\cite{SM_TamascelliPRL2019,SM_SomozaPRL2019,SM_SomozaCommunPhys2023,SM_LorenzoniPRL2024,SM_Caycedo2022}, including hierarchical equations of motion (HEOM) and thermalized time-evolving density operator with orthogonal polynomials algorithm (T-TEDOPA).

\section{Additional text}

\subsection{Phenomenological conditions for long-lived excitonic coherences}

Here we discuss the conditions for long-lived excitonic coherences observed in numerically exact simulations of the FMO complex.

The interaction between exciton states and underdamped intra-pigment vibrational modes is decomposed into two parts $H_{e-v}= H_{e-v}^{(d)} + H_{e-v}^{(o)}$, where $H_{e-v}^{(d)}$ denotes the diagonal terms in the exciton basis
\begin{equation}
    H_{e-v}^{(d)} =\sum_{i=1}^7 |E_i\rangle\langle E_i| \sum_{n=1}^{7}\sum_{k=1}^{62}|\braket{E_i|\epsilon_n}|^2\omega_k\sqrt{s_k}(b_{n,k}+b_{n,k}^{\dagger}),
\end{equation}
and $H_{e-v}^{(o)}$ represents the off-diagonal terms in the exciton basis
\begin{equation}
    H_{e-v}^{(o)} =\sum_{i\neq j}^7 |E_i\rangle\langle E_j| \sum_{n=1}^{7}\sum_{k=1}^{62}\braket{E_i|\epsilon_n}\braket{\epsilon_n|E_j}\omega_k\sqrt{s_k}(b_{n,k}+b_{n,k}^{\dagger}).
\end{equation}
The off-diagonal vibronic couplings $H_{e-v}^{(o)}$ induce transitions between different exciton states while creating or annihilating vibrational excitations. Roughly speaking, if an exciton state $\ket{E_j,0}$ is initially populated with $\ket{0}$ denoting the global vibrational ground state of the 62 intra-pigment modes, the off-diagonal vibronic couplings $H_{e-v}^{(o)}$ induce a transition to $\ket{\psi(t)}=\alpha_0(t)\ket{E_j,0}+\sum_{k=1}^{62}\alpha_k(t)\ket{E_i,1_k}$ over time $t$, where $\ket{1_k}$ denotes a singly-excited vibrational state of the $k$-th intra-pigment mode while all other modes remain in their vibrational ground states. This results in vibronically-generated excitonic coherences given by ${\rm Tr}_v[\langle E_i|\psi(t)\rangle\langle \psi(t)| E_j\rangle]$ where ${\rm Tr}_v$ denotes the partial trace over vibrational degrees of freedom. Since the global vibrational ground state of $\ket{E_j,0}$ and the singly-excited vibrational states of $\ket{E_i,1_k}$ are defined with respect to different vibrational potential surfaces conditioned on the electronic states $\ket{E_j}$ or $\ket{E_i}$, described by the diagonal vibronic couplings $H_{e-v}^{(d)}$, the overlaps $\langle 0|1_k\rangle$ between these vibrational states do not vanish, leading to non-zero excitonic coherence amplitudes. The strength of the off-diagonal vibronic coupling responsible for exciton state transitions $\ket{E_i}\leftrightarrow\ket{E_j}$ depends on the spatial overlap $\beta_{i,j}=\sum_{n=1}^{7}|\braket{E_i|\epsilon_n}\braket{\epsilon_n|E_j}|$ between the exciton states.

\begin{figure}[t]
\includegraphics[width=1.0\textwidth]{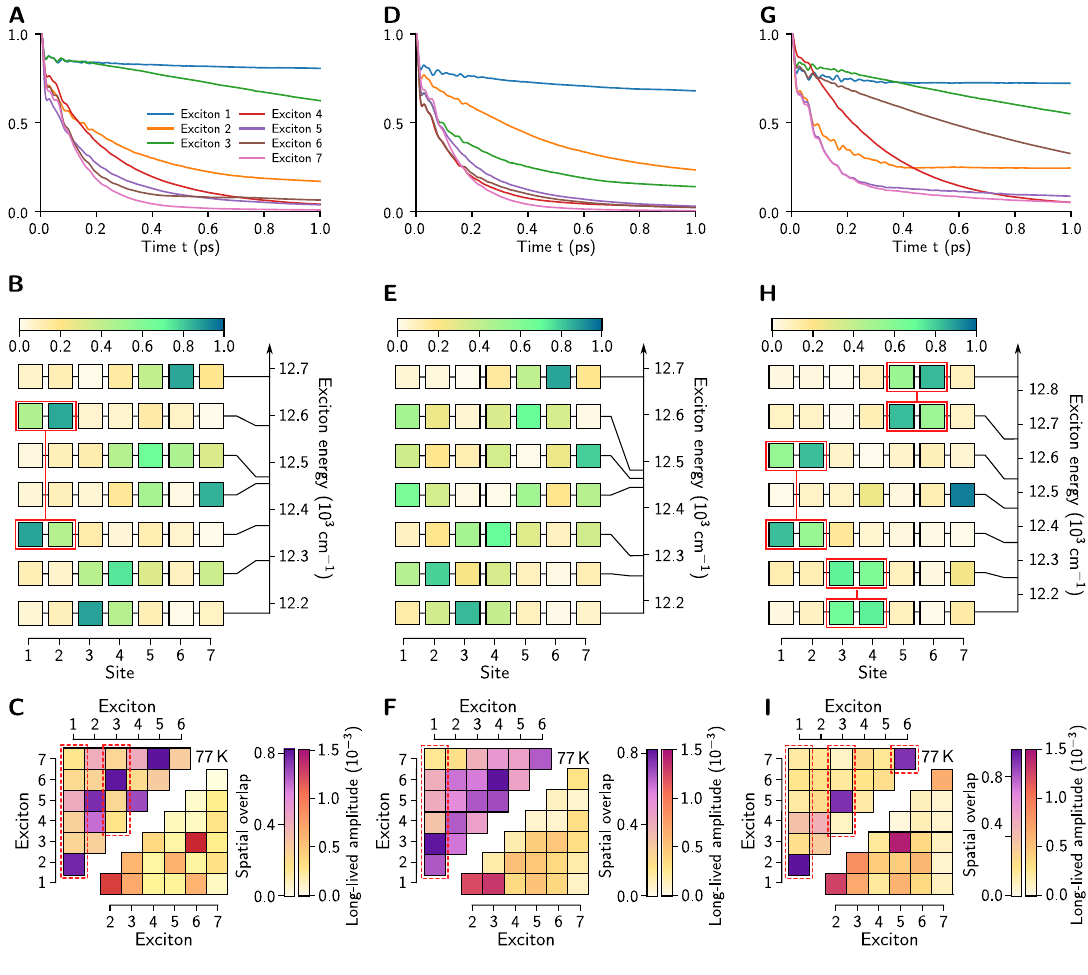}
\caption{\textbf{Exciton population dynamics and phenomenological conditions for long-lived excitonic coherences}. (A) Population dynamics of exciton state $\ket{E_k}$ when initially populated, i.e. $\rho(0)=\ket{E_k}\bra{E_k}$, computed using the electronic parameters estimated in Ref.~\cite{SM_RengerBJ2006} (see Table~\ref{Table_electronic_parameters}). (B) Population distributions of the seven exciton states $\ket{E_k}$ in the site basis. (C) Spatial overlaps $\beta_{i,j}$ between exciton states (purple-yellow scale) and the amplitudes of long-lived excitonic coherences (red-yellow scale). (D-F) Results obtained using the first set of site energies from Table~\ref{Table_fictitious}, where no quasi-dimeric units are formed. (G-I) Results obtained using the second set of site energies from Table~\ref{Table_fictitious}, where three quasi-dimeric units are formed. In all simulations, the FMO phonon spectral density at $T=77\,{\rm K}$ was considered.}
\label{FigS4}
\end{figure}

\begin{table}
\caption{Modified site energies of the FMO complex, corresponding to cases where no quasi-dimeric units or three quasi-dimeric units are formed (see Figs.~\ref{FigS4}D-F and G-I, respectively).}
\label{Table_fictitious}
\begin{ruledtabular}
\begin{tabular}{rrrrrrrr}
$\sf{\epsilon_n (cm^{-1})}$ & 1 & 2 & 3 & 4 & 5 & 6 & 7\\
\hline\hline
No quasi-dimeric units & 12410 & 12290 & 12210 & 12320 & 12480 & 12630 & 12440\\
\hline
Three quasi-dimeric units & 12410 & 12480 & 12210 & 12220 & 12700 & 12780 & 12440 \\
\end{tabular}
\end{ruledtabular}
\end{table}

In case of the FMO complex, which supports more than two exciton states, the previous explanation is insufficient to describe the vibronically-generated excitonic coherence dynamics. If an exciton state $\ket{E_j,0}$ is initially populated and multiple transitions occur consecutively from $\ket{E_j}$ to $\ket{E_i}$, and then to $\ket{E_f}$, the total system at time $t$ may be described by $\ket{\psi(t)}=\alpha_0(t)\ket{E_j,0}+\sum_{k=1}^{62}\alpha_k(t)\ket{E_i,1_k}+\sum_{k=1}^{62}\alpha_k'(t)\ket{E_{f},2_k}$. As the amplitudes $\alpha_k(t)$ responsible for excitonic coherences between $\ket{E_i}$ and $\ket{E_j}$ decrease due to further transitions from $\ket{E_i}$ to $\ket{E_f}$, this effectively induces the dephasing of excitonic coherence $\bra{E_i}\rho(t)\ket{E_j}$ via a relaxation process. Therefore, long-lived oscillations of dynamically-generated excitonic coherences are expected when at least one of the exciton states, $\ket{E_i}$ or $\ket{E_j}$, is either the lowest-energy exciton state $\ket{E_1}$ or a meta-stable state.

For the electronic parameters of the FMO complex estimated in Ref.~\cite{SM_RengerBJ2006} and summarized in Table~\ref{Table_electronic_parameters}, Fig.~\ref{FigS4}A presents the population dynamics of each exciton state $\ket{E_k}$ when initially populated (i.e. $\rho(0)=\ket{E_k}\bra{E_k}$). Notably, the populations of the lowest-energy exciton $\ket{E_1}$ and the exciton state $\ket{E_3}$, which is an electronic eigenstate localized on a quasi-dimeric unit of sites 1 and 2 with a lower exciton energy $E_3<E_6$ (see Fig.~\ref{FigS4}B), remain high for up to $1\,{\rm ps}$, compared to other exciton states. This suggests that $\ket{E_3}$ is a meta-stable state. Fig.~\ref{FigS4}C shows that relatively large spatial overlaps $\beta_{i,j}$ occur between multiple exciton pairs $(i,j)$ (see purple-yellow scale), but only the $(1,2)$ and $(3,6)$ pairs involve either the lowest-energy exciton $\ket{E_1}$ or the meta-stable state $\ket{E_3}$. For these pairs, long-lived excitonic coherences exhibit relatively large amplitudes (see red-yellow scale).

In Figs.~\ref{FigS4}D-F, we modified site energies so that no quasi-dimeric units are formed in the FMO complex using the parameters summarized in the first row of Table~\ref{Table_fictitious}. In this case, we found that there is no meta-stable state, and as a result, only the $(1,2)$ and $(1,3)$ pairs exhibit relatively large overlaps $\beta_{i,j}$ and involve the lowest-energy exciton state $\ket{E_1}$. These pairs exhibit relatively large amplitudes of long-lived excitonic coherences. Similarly, in Figs.~\ref{FigS4}G-I, we considered a different set of site energies that introduce three quasi-dimeric units in the FMO complex (see the second row of Table~\ref{Table_fictitious}). Here, the exciton state $\ket{E_6}$ becomes more stable and has a larger spatial overlap $\beta_{6,7}$ with the highest-energy exciton state $\ket{E_7}$, leading to an enhanced amplitude of the long-lived excitonic coherence between $\ket{E_6}$ and $\ket{E_7}$.

\subsection{Amplitude maps of long-lived oscillations from rephasing ground-state bleaching and stimulated emission signals}

\begin{figure*}[b]
\includegraphics[width=0.66\textwidth]{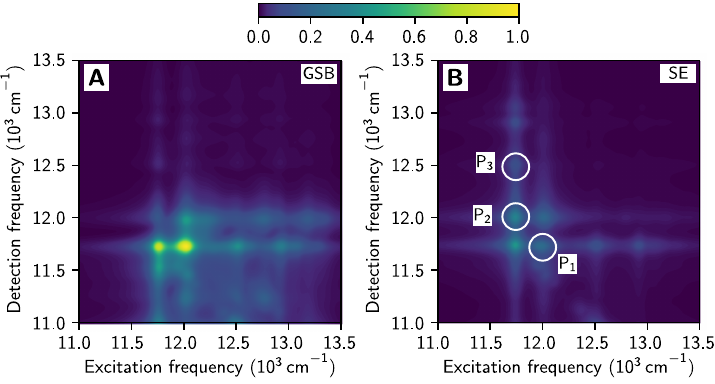}
\caption{\textbf{Amplitude maps of long-lived oscillatory rephasing signals}. The absolute value of the real part of the oscillatory (A) GSB and (B) SE signals, obtained via a high-pass filter, was integrated over the waiting time $t_2\in [0.2,1.0]\,{\rm ps}$ for each spectral position $(\omega_1,\omega_3)$. A dimeric PPC model ($\epsilon_1-\epsilon_2 = 200\,{\rm cm}^{-1}$, $V_{12}=100\,{\rm cm}^{-1}$, orthogonal transition dipole moments, $\bm{\mu}_1\cdot \bm{\mu}_2=0$, with identical magnitudes, $\bm{\mu}_1\cdot \bm{\mu}_1=\bm{\mu}_2\cdot \bm{\mu}_2$) was considered with the FMO phonon spectral density at $77\,{\rm K}$.}
\label{FigS5}
\end{figure*}

For a given pair of excitation and detection frequencies $(\omega_1,\omega_3)$, we decomposed the rephasing 2D signals along the waiting time $t_2$ into damped oscillations and exponential decay terms. The non-oscillatory exponential components were removed using a high-pass filter with a frequency cutoff $100\,{\rm cm}^{-1}$, which is lower than the excitonic splitting $\Delta_{12}\approx 283\,{\rm cm}^{-1}$ of the dimeric PPC model used in 2D simulations, as well as the vibrational frequencies of the intra-pigment modes summarized in Table~\ref{Table_intrapigment}. To identify the spectral regions $(\omega_1,\omega_3)$ where the amplitudes of long-lived oscillations are maximized, we applied the high-pass filter to ground-state bleaching (GSB) and stimulated emission (SE) signals, separately. We then integrated the absolute value of the high-pass filtered signals over a time window $t_2\in [0.2,1.0]\,{\rm ps}$, as shown in Figs.~\ref{FigS5}A and B. The three peak positions, P1, P2 and P3, in Fig.~\ref{FigS5}B of the rephasing SE spectra were considered in Fig.~5B of the main text.

\end{widetext}

\begin{thebibliography}{99}

\bibitem{BlankenshipBook2002} R. E. Blankenship, Molecular mechanisms of photosynthesis (Wiley-Blackwell, 2002).

\bibitem{VultoJPCB1998} S. I. E. Vulto, M. A. de Baat, R. J. W. Louwe, H. P. Permentier, T. Neef, M. Miller, H. van Amerongen, T. J. Aartsma, Exciton simulations of optical spectra of the FMO complex from the green sulfur bacterium Chlorobium Tepidum at 6 K. J. Phys. Chem. B. {\bf 102}, 9577 (1998).

\bibitem{RengerBJ2006} J. Adolphs, T. Renger, How proteins trigger excitation energy transfer in the FMO complex of green sulfur bacteria. Biophys. J. {\bf 91}, 2778-2797 (2006).

\bibitem{BuschJPCL2011} M. S. am Busch, F. M{\"u}h, M. E.-A. Madjet, T. Renger, The eighth bacteriochlorophyll completes the excitation energy funnel in the FMO protein. J. Phys. Chem. Lett. {\bf 2}, 93–98 (2011).

\bibitem{RengerJPCB2012} T. Renger, A. Klinger, F. Steinecker, M. S. am Busch, J. Numata, F. M{\"u}h, Normal mode analysis of the spectral density of the Fenna–Matthews–Olson light-harvesting protein: how the protein dissipates the excess energy of excitons. J. Phys. Chem. B {\bf 116}, 14565–14580 (2012).

\bibitem{CokerJPCL2016} M. K. Lee, D. F. Coker, Modeling electronic-nuclear interactions for excitation energy transfer processes in light-harvesting complexes. J. Phys. Chem. Lett. {\bf 7}, 3171–3178 (2016).

\bibitem{RheePCCP2018} C. W. Kim, B. Choi, Y. M. Rhee, Excited state energy fluctuations in the Fenna–Matthews–Olson complex from molecular dynamics simulations with interpolated chromophore potentials. Phys. Chem. Chem. Phys. {\bf 20}, 3310-3319 (2018).

\bibitem{KleinekathoferPR2023} S. Maity, U. Kleinekathöfer, Recent progress in atomistic modeling of light-harvesting complexes: a mini review. Photosynth. Res. {\bf 156}, 147–162 (2023).

\bibitem{RatsepJL2007} M. R{\"a}tsep, A. Freiberg, Electron phonon and vibronic couplings in the FMO bacteriochlorophyll a antenna complex studied by difference fluorescence line narrowing. J. Lumin. {\bf 127}, 251-259 (2007).

\bibitem{Caycedo2022} F. Caycedo-Soler, A. Mattioni, J. Lim, T. Renger, S. F. Huelga, M. B. Plenio, Exact simulation of pigment-protein complexes unveils vibronic renormalization of electronic parameters in ultrafast spectroscopy. Nat. Commun. {\bf 13}, 2912 (2022).

\bibitem{StrathearnNC2018} A. Strathearn, P. Kirton, D. Kilda, J. Keeling, B. W. Lovett, Efficient non-Markovian quantum dynamics using time-evolving matrix product operators. Nat. Commun. {\bf 9}, 3322 (2018).

\bibitem{TamascelliPRL2019} D. Tamascelli, A. Smirne, J. Lim, S. F. Huelga, and M. B. Plenio, Efficient simulation of finite-temperature open quantum systems. Phys. Rev. Lett. {\bf 123}, 090402 (2019).

\bibitem{SomozaPRL2019} A. D. Somoza, O. Marty, J. Lim, S. F. Huelga, M. B. Plenio, Dissipation-assisted matrix product factorization. Phys. Rev. Lett. {\bf 123}, 100502 (2019).

\bibitem{CygorekNP2022} M. Cygorek, M. Cosacchi, A. Vagov, V. M. Axt, B. W. Lovett, J. Keeling, E. M. Gauger, Simulation of open quantum systems by automated compression of arbitrary environments. Nat. Phys. {\bf 18}, 662–668 (2022).

\bibitem{LorenzoniPRL2024}  N. Lorenzoni, N. Cho, J. Lim, D. Tamascelli, S. F. Huelga, M. B. Plenio, Systematic coarse graining of environments for the nonperturbative simulation of open quantum systems. Phys. Rev. Lett. {\bf 132}, 100403 (2024).

\bibitem{IshizakiPNAS2009} A. Ishizaki, G. R. Fleming, Theoretical examination of quantum coherence in a photosynthetic system at physiological temperature. Proc. Natl Acad. Sci. USA {\bf 106}, 17255-17260 (2009).

\bibitem{NalbachPRE2011} P. Nalbach, D. Braun, M. Thorwart, Exciton transfer dynamics and quantumness of energy transfer in the Fenna-Matthews-Olson complex. Phys. Rev. E {\bf 84}, 041926 (2011).

\bibitem{KreisbeckJPCL2012} C. Kreisbeck, T. Kramer, Long-lived electronic coherence in dissipative exciton dynamics of light-harvesting complexes. J. Phys. Chem. Lett. {\bf 3}, 2828–2833 (2012).

\bibitem{BlauPNAS2018} S. M. Blau, D. I. G. Bennett, C. Kreisbeck, G. D. Scholes, A. Aspuru-Guzik, Local protein solvation drives direct down-conversion in phycobiliprotein PC645 via incoherent vibronic transport. Proc. Natl Acad. Sci. USA {\bf 115}, E3342-E3350 (2018).

\bibitem{JonasARPC2003} D. M. Jonas, Two-dimensional femtosecond spectroscopy. Annu. Rev. Phys. Chem. {\bf 54}, 425-463 (2003).

\bibitem{BrixnerJCP2004} T. Brixner, T. Man{\v c}al, I. V. Stiopkin, G. R. Fleming, Phase-stabilized two-dimensional electronic spectroscopy. J. Chem. Phys. {\bf 121}, 4221–4236 (2004).

\bibitem{ButkusCPL2012} V. Butkus, D. Zigmantas, L. Valkunas, D. Abramavicius, Vibrational vs. electronic coherences in 2D spectrum of molecular systems. Chem. Phys. Lett. {\bf 545}, 40-43 (2012).

\bibitem{TiwariPNAS2013} V. Tiwari, W. K. Peters, D. M. Jonas, Electronic resonance with anticorrelated pigment vibrations drives photosynthetic energy transfer outside the adiabatic framework. Proc. Natl Acad. Sci. USA {\bf 110}, 1203–1208 (2013).

\bibitem{PlenioJCP2013} M. B. Plenio, J. Almeida, S. F. Huelga, Origin of long-lived oscillations in 2D-spectra of a quantum vibronic model: electronic versus vibrational coherence. J. Chem. Phys. {\bf 139}, 235102 (2013).

\bibitem{DuanPNAS2017} H.-G. Duan, V. I. Prokhorenko, R. J. Cogdell, K. Ashraf, A. L. Stevens, M. Thorwart, R. J. D. Miller, Nature does not rely on long-lived electronic quantum coherence for photosynthetic energy transfer. Proc. Natl Acad. Sci. USA {\bf 114}, 8493-8498 (2017).

\bibitem{ThyrhaugNC2018} E. Thyrhaug, R. Tempelaar, M. J. P. Alcocer, K. {\v Z}{\' i}dek, D. B{\' i}na, J. Knoester, T. L. C. Jansen, D. Zigmantas, Identification and characterization of diverse coherences in the Fenna–Matthews–Olson complex. Nat. Chem. {\bf 10}, 780–786 (2018).

\bibitem{EngelNat2007} G. S. Engel, T. R. Calhoun, E. L. Read, T. Ahn, T. Man{\v c}al, Y. Cheng, R. E. Blankenship, G. R. Fleming, Evidence for wavelike energy transfer through quantum coherence in photosynthetic systems. Nature {\bf 446}, 782–786 (2007).

\bibitem{EngelPNAS2010} G. Panitchayangkoon, D. Hayes, K. A. Fransted, J. R. Caram, E. Harel, J. Wen, R. E. Blankenship, G. S. Engel, Long-lived quantum coherence in photosynthetic complexes at physiological temperature. Proc. Natl Acad. Sci. USA {\bf 107}, 12766-12770 (2010).

\bibitem{CaoSA2020} J. Cao, R. J. Cogdell, D. F. Coker, H.-G. Duan, J. Hauer, U. Kleinekath{\"o}fer, T. L. C. Jansen, T. Man{\v c}al, R. J. D. Miller, J. P. Ogilvie, V. I. Prokhorenko, T. Renger, H.-S. Tan, R. Tempelaar, M. Thorwart, E. Thyrhaug, S. Westenhoff, D. Zigmantas, Quantum biology revisited. Sci. Adv. {\bf 6}, eaaz4888 (2020).

\bibitem{MauriNatChem2018} M. Maiuri, E. E. Ostroumov, R. G. Saer, R. E. Blankenship, and G. D. Scholes, Coherent wavepackets in the Fenna–Matthews–Olson complex are robust to excitonic-structure perturbations caused by mutagenesis. Nat. Chem. {\bf 10}, 177–183 (2018).

\bibitem{Petruccione2002} H. Breuer, F. Petruccione, The theory of open quantum systems (Oxford University Press, 2002).

\bibitem{KolliJCP2012} A. Kolli, E. J. O’Reilly, G. D. Scholes, A. Olaya-Castro, The fundamental role of quantized vibrations in coherent light harvesting by cryptophyte algae. J. Chem. Phys. {\bf 137}, 174109 (2012).

\bibitem{ChinNP2013} A. W. Chin, J. Prior, R. Rosenbach, F. Caycedo-Soler, S. F. Huelga, M. B. Plenio, The role of non-equilibrium vibrational structures in electronic coherence and recoherence in pigment–protein complexes. Nat. Phys. {\bf 9}, 113-118 (2013).

\bibitem{ChoBook2009} M. Cho, Two-dimensional optical spectroscopy (CRC Press, 2009).

\bibitem{Lim2019} J. Lim, C. M. B{\"o}sen, A. D. Somoza, C. P. Koch, M. B. Plenio, S. F. Huelga, Multi-color quantum control for suppressing ground state coherences in two-dimensional electronic spectroscopy. Phys. Rev. Lett. {\bf 123}, 233201 (2019).

\bibitem{ZazubovichJPCB2001} V. Zazubovich, I. Tibe, G. J. Small, Bacteriochlorophyll a Franck-Condon factors for the $S_0\rightarrow S_1(Q_y)$ transition. J. Phys. Chem. B {\bf 105}, 12410-12417 (2001).

\bibitem{OgilvieJPCL2018} V. R. Policht, A. Niedringhaus, J. P. Ogilvie, Characterization of vibrational coherence in monomeric bacteriochlorophyll a by two-dimensional electronic spectroscopy. J. Phys. Chem. Lett. {\bf 9}, 6631–6637 (2018).

\bibitem{HarelJPCL2018} S. Irgen-Gioro, A. P. Spencer, W. O. Hutson, E. Harel, Coherences of bacteriochlorophyll a uncovered using 3D-electronic spectroscopy. J. Phys. Chem. Lett. {\bf 9}, 6077–6081 (2018).

\bibitem{ColliniCP2019} E. Meneghin, D. Pedron, E. Collini, Characterization of the coherent dynamics of bacteriochlorophyll a in solution. Chem. Phys. {\bf 519}, 85-91 (2019).

\bibitem{SenlikJPCL2015} S. S. Senlik, V. R. Policht, J. P. Ogilvie, Two-color nonlinear spectroscopy for the rapid acquisition of coherent dynamics. J. Phys. Chem. Lett. 6, 2413–2420 (2015).

\bibitem{BermanNucleicAcidsRes2000} H. M. Berman, J. Westbrook, Z. Feng, G. Gilliland, T. N. Bhat, H. Weissig, I. N. Shindyalov, P. E. Bourne, The protein data bank. Nucleic Acids Res. {\bf 28}, 235–242 (2000).  https://www.rcsb.org/.

\bibitem{TronrudPhotosynthRes2009} D. E. Tronrud, J. Wen, L. Gay, R. E. Blankenship, The structural basis for the difference in absorbance spectra for the FMO antenna protein from various green sulfur bacteria. Photosynth. Res. {\bf 100}, 79-87 (2009).

\bibitem{FMOstructure} D. E. Tronrud, A. Camara-Artigas, R. E. Blankenship, J. P. Allen, Crystal structure of the Fenna-Matthews-Olson Protein from Chlorobaculum Tepidum (2009). https://doi.org/10.2210/pdb3ENI/pdb.

\bibitem{SehnalNucleicAcidsRes2021} D. Sehnal, S. Bittrich, M. Deshpande, R. Svobodov{\' a}, K. Berka, V. Bazgier, S. Velankar, S. K. Burley, J. Ko{\v c}a, A. S. Rose, Mol* viewer: modern web app for 3D visualization and analysis of large biomolecular structures. Nucleic Acids Res. {\bf 49}, W431-W437 (2021).

\end{thebibliography}

\begin{thebibliography}{99}
\bibitem{SM_RengerBJ2006} J. Adolphs, T. Renger, How proteins trigger excitation energy transfer in the FMO complex of green sulfur bacteria. Biophys. J. {\bf 91}, 2778-2797 (2006).

\bibitem{SM_RatsepJL2007} M. R{\"a}tsep, A. Freiberg, Electron phonon and vibronic couplings in the FMO bacteriochlorophyll a antenna complex studied by difference fluorescence line narrowing. J. Lumin. {\bf 127}, 251-259 (2007).

\bibitem{SM_DuanPNAS2017} H.-G. Duan, V. I. Prokhorenko, R. J. Cogdell, K. Ashraf, A. L. Stevens, M. Thorwart, R. J. D. Miller, Nature does not rely on long-lived electronic quantum coherence for photosynthetic energy transfer. Proc. Natl Acad. Sci. USA {\bf 114}, 8493-8498 (2017).

\bibitem{SM_SomozaPRL2019} A. D. Somoza, O. Marty, J. Lim, S. F. Huelga, M. B. Plenio, Dissipation-assisted matrix product factorization. Phys. Rev. Lett. {\bf 123}, 100502 (2019).


\bibitem{SM_MelkozernovPhotosynthRes1998} A. N. Melkozernov, J. M. Olson, Y. Li, J. P. Allen, and R. E. Blankenship, Orientation and excitonic interactions of the Fenna–Matthews–Olson bacteriochlorophyll a protein in membranes of the green sulfur bacterium Chlorobium tepidum. Photosynth. Res. {\bf 56}, 315 (1998).


\bibitem{SM_SomozaCommunPhys2023} A. D. Somoza, N. Lorenzoni, J. Lim, S. F. Huelga, and M. B. Plenio, Driving force and nonequilibrium vibronic dynamics in charge separation of strongly bound electron-hole pairs. Commun. Phys. {\bf 6}, 65 (2023).

\bibitem{SM_LorenzoniPRL2024}  N. Lorenzoni, N. Cho, J. Lim, D. Tamascelli, S. F. Huelga, M. B. Plenio, Systematic coarse graining of environments for the nonperturbative simulation of open quantum systems. Phys. Rev. Lett. {\bf 132}, 100403 (2024).

\bibitem{SM_TamascelliPRL2019} D. Tamascelli, A. Smirne, J. Lim, S. F. Huelga, and M. B. Plenio, Efficient simulation of finite-temperature open quantum systems. Phys. Rev. Lett. {\bf 123}, 090402 (2019).

\bibitem{SM_TamascelliPRL2018} D. Tamascelli, A. Smirne, S. F. Huelga, and M. B. Plenio, Nonperturbative treatment of non-Markovian dynamics of open quantum systems. Phys. Rev. Lett. {\bf 120}, 030402 (2018).

\bibitem{SM_Caycedo2022} F. Caycedo-Soler, A. Mattioni, J. Lim, T. Renger, S. F. Huelga, M. B. Plenio, Exact simulation of pigment-protein complexes unveils vibronic renormalization of electronic parameters in ultrafast spectroscopy. Nat. Commun. {\bf 13}, 2912 (2022).

\end{thebibliography}
\end{document}